\documentclass[aps,twocolumn,superscriptaddress,reprint,
nofootinbib,tightenlines]{revtex4-1}

\newcommand{\eqref}[1]{(\ref{#1})\xspace}

\usepackage{color}
\usepackage{graphicx} 
\usepackage{bm}       
\usepackage{amssymb}
\usepackage{xspace}                    

\newcommand{\de}{\partial}
\newcommand{\half}{\frac{1}{2}}

\newcommand{\HIGS}{HI$\gamma$S\xspace}
\newcommand{\threeHe}{${}^3$He\xspace}
\newcommand{\fourHe}{${}^4$He\xspace}

\newcommand{\alphae}{\ensuremath{\alpha_{E1}}}
\newcommand{\betam}{\ensuremath{\beta_{M1}}}
\newcommand{\gammaee}{\ensuremath{\gamma_{E1E1}}}

\newcommand{\gammaem}{\ensuremath{\gamma_{E1M2}}}
\newcommand{\gammame}{\ensuremath{\gamma_{M1E2}}}

\newcommand{\alphaep}{\ensuremath{\alpha_{E1}^{(\mathrm{p})}}}
\newcommand{\betamp}{\ensuremath{\beta_{M1}^{(\mathrm{p})}}}
\newcommand{\alphaen}{\ensuremath{\alpha_{E1}^{(\mathrm{n})}}}
\newcommand{\betamn}{\ensuremath{\beta_{M1}^{(\mathrm{n})}}}

\newcommand{\mpi}{\ensuremath{m_\pi}}     
\newcommand{\MeV}{\ensuremath{\mathrm{MeV}}}
\newcommand{\fm}{\ensuremath{\mathrm{fm}}}
\newcommand{\ChiEFT}{$\chi$EFT\xspace}

\newcommand{\NXLO}[1]{N\ensuremath{{}^{#1}}LO\xspace}

\newcommand{\calL}{\mathcal{L}}
\newcommand{\calO}{\mathcal{O}}

\begin{document}

\title{Compton Scattering and Nucleon Polarisabilities\\ in Chiral EFT: 
Status and Future}

\author{Harald W.~Grie\3hammer} \affiliation{Institute for Nuclear Studies,
  Department of Physics, George Washington University, Washington DC 20052,
  USA} \email{hgrie@gwu.edu; corresponding author. Invited Contribution to the
  \textsc{ 22nd International Spin Symposium} (SPIN 2016), University of
  Illinois, Urbana (USA), 26-30 September 2016.}

\author{Judith A.~McGovern} \affiliation{School of Physics and Astronomy, The
  University of Manchester, Manchester M13 9PL, UK} \author{Daniel
  R.~Phillips} \affiliation{Dept.~of Physics and Astronomy, Institute of
  Nuclear and Particle Physics, Ohio University, Athens OH 45701, USA}

\begin{abstract}
  We review theoretical progress and prospects for determining the nucleon's
  static dipole polarisabilities from Compton scattering on few-nucleon
  targets, including new values and an emphasis on what polarised targets and
  beams can contribute; see Refs.~\cite{Gr12, McGovern:2012ew,
    Griesshammer:2013vga, Myers:2014ace, latticepols} for details and a more
  thorough bibliography.
\end{abstract}

\maketitle

\section{Why Compton Scattering?}
\label{sec:intro}

Let us start with an even simpler question: Why can you see any speaker at a
conference? Light shines on matter, gets absorbed, and is re-emitted before it
reaches your eyes. That is Compton scattering $\gamma \mathrm{X}\to\gamma
\mathrm{X}$. A white shirt and red jumper reflect light differently because they
have different chemical compositions: one re-emits radiation quite uniformly
over the visible band, the other absorbs most non-red photons. The speaker's
attire therefore not only shows their fashion sense (or lack thereof), but
betrays information about the stuff of which they are made.

Let's be a bit more scientific. 
In Compton scattering $\gamma \mathrm{X}\to\gamma \mathrm{X}$, t%
he electromagnetic field of a real photon induces radiation multipoles by
displacing 
charged constituents and currents inside the target.  The energy- and
angle-dependence of the emitted radiation carries information on the
interactions of the constituents. In Hadronic Physics, it elucidates the
distribution, symmetries and dynamics of the charges and currents which
constitute the low-energy degrees of freedom inside the nucleon and nucleus,
and -- for nuclei 
-- the interactions between nucleons,
complementing information from one-photon data like form factors; see e.g.~a
recent review~\cite{Gr12}.  In contradistinction to many other electromagnetic
processes, such structure effects have only recently been subjected to a
multipole-analysis.  The Fourier transforms of the corresponding temporal
response functions are the proportionality constants between incident field
and induced multipole. These \emph{energy-dependent polarisabilities}
parametrise the stiffness of the nucleon $N$ (spin $\vec{\sigma}/2$)
against transitions ${Xl\to Yl^\prime}$ of given photon multipolarity at fixed
frequency $\omega$ ($l^\prime=l\pm\{0;1\}$; $X,Y=E,M$; $T_{ij}=\half (\de_iT_j
+ \de_jT_i)$; $T=E,B$). Up to about $400\;\MeV$, the relevant terms are:
\begin{widetext}
\begin{equation}
\begin{array}{ll}
  \label{polsfromints}
  \calL_\mathrm{pol}=2\pi\;N^\dagger
  \;\big[&{\alpha_{E1}(\omega)}\;\vec{E}^2\;+
  \;{\beta_{M1}(\omega)}\;\vec{B}^2\; 
 +\;{\gamma_{E1E1}(\omega)}
  \;\vec{\sigma}\cdot(\vec{E}\times\dot{\vec{E}})\;
  +\;{\gamma_{M1M1}(\omega)}
  \;\vec{\sigma}\cdot(\vec{B}\times\dot{\vec{B}})\\[0.5ex]&
  -\;2{\gamma_{M1E2}(\omega)}\;\sigma^i\;B^j\;E_{ij}\;+
  \;2{\gamma_{E1M2}(\omega)}\;\sigma^i\;E^j\;B_{ij} \;
+\;\dots\;(\mbox{photon
    multipoles beyond dipole}) \big]\;N
\end{array}
\end{equation} 
\end{widetext}
The two spin-independent polarisabilities $\alpha_{E1}(\omega)$ and
$\beta_{M1}(\omega)$ parametrise electric and magnetic dipole transitions.  Of
particular interest at present are now the four dipole spin-polarisabilities
$\gamma_{E1E1}(\omega)$, $\gamma_{M1M1}(\omega)$, $\gamma_{E1M2}(\omega)$ and
$\gamma_{M1E2}(\omega)$. They encode the response of the nucleon's spin
structure, i.e.~of the spin constituents, and complement JLab experiments at
much higher energies.  Intuitively interpreted, the electromagnetic field
associated with the spin degrees causes birefringence in the
nucleon 
(cf.~classical Faraday-effect). Only the linear combinations $\gamma_0$ and
$\gamma_\pi$ of scattering under ($0^\circ$ and $180^\circ$ scattering) are
somewhat constrained by data or phenomenology , with conflicting results for
the proton (MAMI, LEGS) and large error-bars for the neutron.

The spin polarisabilities are of particular interest since they provide an
excellent window into the photon-pion-nucleon system, i.e.~into the charged
pion cloud around the nucleon.  While the $\gamma\pi$ interactions do not
depend on the nucleon spin, the orientation of the pion cloud itself depends
on the nucleon spin. Indeed, recall that the dominant $\pi N$ and $\gamma\pi
N$ interactions are related by minimal substitution, $\calL_{\gamma\pi
  N}=-\frac{g_A}{2f_\pi}\vec{\sigma}\cdot(\vec{q}+e\,\vec{\epsilon})$, where
$\vec{q}$ is the pion momentum and $\vec{\epsilon}$ the photon
polarisation. So both photon and pion emission or absorption is strongly
dependent on the nucleon spin. As this picture persists at higher orders of
the chiral expansion, Compton scattering with polarised photons on polarised
nucleons provides stringent tests of \ChiEFT -- and the spin polarisabilities
are just the observables to look at.

Since the polarisabilities are the parameters of a multipole decomposition,
they do not contain more information than the full amplitudes, but
characteristic signatures in specific channels are easier to interpret. For
example, the strong $\omega$-dependence of $\beta_{M1}(\omega)$ and
$\gamma_{M1M1}(\omega)$ for $\omega\gtrsim100\;\MeV$ comes from the strong
para-magnetic $\gamma \mathrm{N}\Delta$ transition.  The $\Delta(1232)$ thus
enters dynamically well below the resonance region.  The electric
polarisability, in turn, exhibits a pronounced cusp at the pion-production
threshold.  As soon as an inelastic channel opens, namely at the
pion-production threshold, the dynamical polarisabilities become complex, and
their imaginary parts are directly related to the pion-photoproduction
multipoles.  Polarisabilities also test our understanding of the subtle
interplay between electromagnetic and strong interactions:
They enter in the two-photon-exchange contribution to the Lamb shift in muonic
hydrogen~\cite{Pohl:2013yb}
. 
%
And the Cottingham Sum rule relates the proton-neutron difference in
$\betam$ to the proton-neutron electromagnetic mass difference. There is a
cantankerous controversy about the precise nature of the
relationship~\cite{WalkerLoud:2012bg,Gasser:2015dwa}, but there is little
doubt that it tests our understanding of the subtle interplay between
electromagnetic and strong interactions in a fundamental observable.
%
Finally, nuclear targets provide an opportunity to study not only neutron
polarisabilities, but indirectly also the nuclear force, since the photons
couple to the charged pion-exchange currents. 

\section{Where We Are}
\label{sec:status}

The values $\alpha_{E1}(\omega=0)$ etc.~are often called ``the (static)
polarisabilities''; they compress the richness of information from data which
is available in a wide range of energies between about $70\;\MeV$ and the
$\Delta$ resonance region, extrapolating it into just a few numbers. In the
canonical units of $10^{-4}\;\fm^3$, the results of our most recent \ChiEFT
extractions are, including an estimate of the residual theoretical
uncertainties from order-by-order convergence (see Fig.~\ref{fig:comptonPDG}
and~\cite{McGovern:2012ew, Myers:2014ace}): 
\begin{equation}
  \label{eq:polsresults}
\begin{array}{l}
  \alphaep=10.65\pm0.35_{\mathrm{stat}}\pm0.2_\mathrm{Baldin}\pm0.3_\mathrm{th}
  \\
  \betamp = 3.15\mp0.35_\mathrm{stat}\pm0.2_\mathrm{Baldin}\mp0.3_\mathrm{th}
  \\[0.5ex]
  \alphaen =11.55\pm 1.25_\mathrm{stat}\pm0.2_\mathrm{Baldin}\pm0.8_\mathrm{th}\\
  \betamn =3.65\mp 1.25_\mathrm{stat}\pm0.2_\mathrm{Baldin}\mp0.8_\mathrm{th}
\end{array}
\end{equation}
For the proton, we checked statistical data consistency, made sure the results
are compatible with the Baldin Sum rule,
fit polarisabilities at $\omega\lesssim170\;\MeV$ and $\Delta(1232)$
parameters above that, iterated until convergence is reached, and finally
find a satisfactory $\chi^2=113.2$ for $135$ degrees of freedom.
The fit quality for the neutron is addressed below.

We also predicted the spin values~\cite{McGovern:2012ew, latticepols}, prior
to the first MAMI data taken at $290\;\MeV$, which provide the first
significant constraints on individual proton spin
polarizabilities~\cite{Martel:2014pba} (in their canonical units of
$10^{-4}\;\fm^4$):
\begin{widetext}
\begin{equation}
  \label{eq:spinpols}
\hspace*{-1ex}
\begin{tabular}{lllll}
  & {$\gamma_{E1E1}$} &{$\gamma_{M1M1}$}
   &{$\gamma_{E1M2}$}&{$\gamma_{M1E2}$}\\[0.5ex]
  \ChiEFT neutron&$-4.0\pm1.9_\mathrm{th}$&
  $1.3\pm0.5_\mathrm{stat}\pm0.6_\mathrm{th}$&
  $-0.1\pm0.6_\mathrm{th}$& $2.4\pm0.5_\mathrm{th}$\\
  \ChiEFT proton&$-1.1\pm1.9_\mathrm{th}$&
  ${2.2\pm0.5_\mathrm{stat}\pm0.6_\mathrm{th}}$ &
  $-0.4\pm0.6_\mathrm{th}$& $1.9\pm0.5_\mathrm{th}$\\[0.5ex]
  proton MAMI&$-3.5\pm1.2$&$3.2\pm0.9$&$-0.7\pm1.2$&
  $2.0\pm0.3$
\end{tabular}
\end{equation}
\vspace*{-3ex}
\begin{figure*}[!thb]
\centering
\parbox{0.47\linewidth}{
  \includegraphics[height=43ex]
  {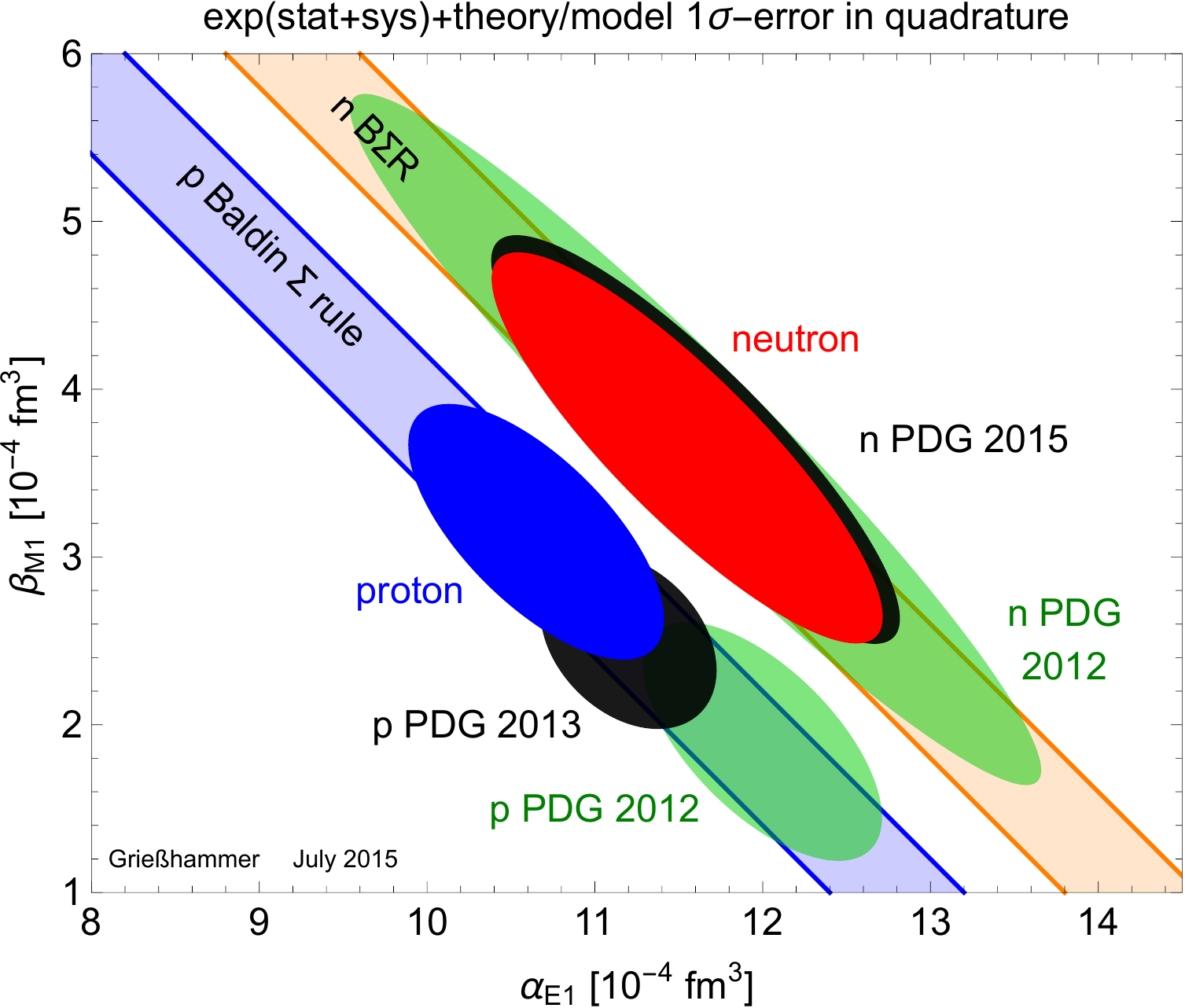}}
\hspace*{4ex}
\parbox{0.47\linewidth}{\includegraphics[height=43ex,clip=,bb = 47 24 471
  328]{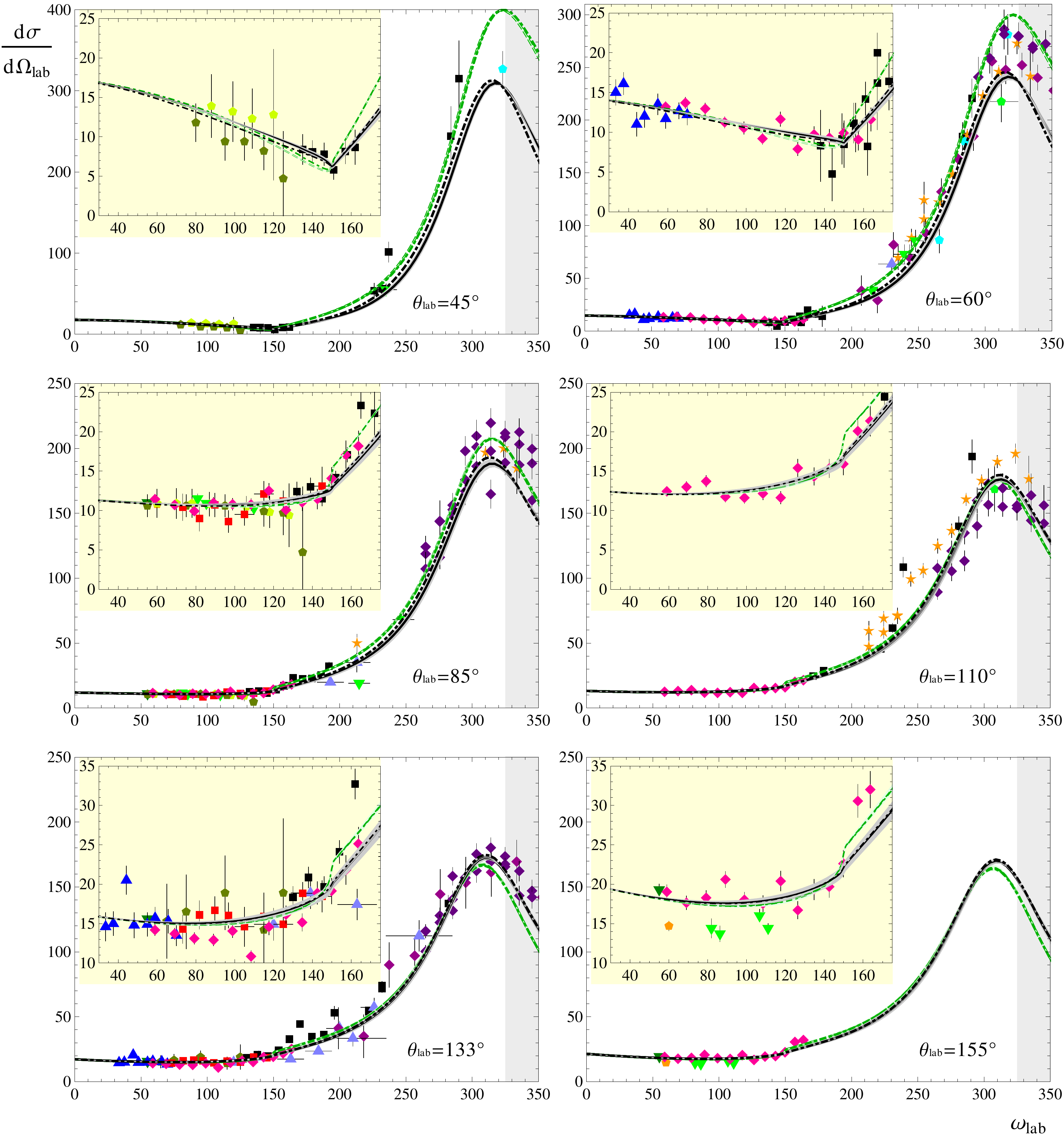} }
\caption{\label{fig:comptonPDG} \emph{Top}: Static scalar polarisabilities in our fits
  (red: proton; blue: neutron); PDG listings prior to (green) and after our
  extractions (black). $1\sigma$ errors, with statistic, systematic and theory
  error added in quadrature. \emph{Bottom}: Example  of data and \ChiEFT
  result, with fit uncertainties, as function of $\omega$; see Ref.~\cite{McGovern:2012ew} for details.}
\end{figure*}
\end{widetext}
A major concerted experimental effort at \HIGS, MAMI and MAXlab will provide
new, high-accuracy data -- including on observables with beam and/or target
polarisations; see e.g.~\cite{Weller:2009zz, Huber:2015uza}.
%
Interpretation of such data of course requires commensurate theory
support. One must carefully evaluate the consistency of the data to reveal
hidden systematic errors; subtract binding effects in few-nucleon systems;
extract the polarisabilities; identify their underlying mechanisms and relate
them to QCD -- and all that with reproducible theoretical uncertainties and
minimal theoretical bias.

Prompted by experimental colleagues 
, theorists with backgrounds in several variants of Dispersion Relations and
Effective Field Theories summarised the present common theoretical
understanding as follows~\cite{theoryletter}. (1) Static polarisabilities can
be extracted from future data well below the pion-production threshold with
high theoretical accuracy and minimal theory error. (2) Data around and above
the pion production threshold show increased sensitivity to the spin
polarisabilities and will help to understand and resolve some discrepancies
between different approaches. (3) All theoretical approaches resort to
well-motivated but not fully controlled approximations around and above the
$\Delta(1232)$ resonance. In the longer term, theorists welcome a complete set
of experiments up to the pion production threshold to disentangle detailed
information from the energy dependence of the Compton multipoles.

\section{Chiral Effective Field Theory}

\ChiEFT, the low-energy theory of QCD and extension of Chiral Perturbation
Theory to few-nucleon systems, has been quite successful in proton and
few-nucleon Compton scattering. Its purely-mesonic sector is Chiral
Perturbation Theory ($\chi$PT); and its one-nucleon sector is Baryon $\chi$PT,
or Heavy-Baryon $\chi$PT, when an additional expansion in the nucleon mass as
a heavy scale is employed which reduces the theory to a non-relativistic one
at leading-order. \ChiEFT generates the most general amplitude consistent with
gauge invariance, the pattern of chiral-symmetry breaking, and Lorentz
covariance. With explicit $\Delta(1232)$ degrees of freedom, its low-energy
scales are the pion mass $\mpi\approx 140\;\MeV$; the Delta-nucleon mass
splitting $\Delta_M\approx290\;\MeV$; and the photon energy $\omega$. When
measured in units of a natural ``high'' scale
$\Lambda
\approx800\;\MeV$ at which this variant can be expected to break down because
new degrees of freedom become dynamical, Pascalutsa and Phillips identified
one common parameter with magnitude smaller than $1$:
$\delta\equiv\frac{\Delta_M}{\Lambda}\approx\left(\frac{m_\pi}{\Lambda}\right)^{1/2}$,
where the latter is a convenient numerical coincidence~\cite{PP03}.  Recently,
we derived single-nucleon Compton amplitudes from zero energy up to about
$350\;\MeV$. For $\omega\lesssim\mpi$, they contain all contributions at
$\calO(e^2\delta^4)$ (\NXLO{4}, accuracy $\delta^5\lesssim2\%$), and for
$\omega\sim \Delta_M$ all at $\calO(e^2\delta^0)$ (NLO, accuracy
$\delta^2\lesssim20\%$)~\cite{Gr12, McGovern:2012ew}.

A reproducible, rigorous and systematically improvable \emph{a priori}
estimate of theoretical accuracies of observables, as in
Eqs.~\eqref{eq:polsresults} and~\eqref{eq:spinpols} or Fig.~\ref{fig:lattice},
is of course essential to uniquely disentangle chiral dynamics from
data. Since the \ChiEFT result is ordered in powers of $\delta<1$, it provides
just that.
Recently, the procedure to justify such estimates was codified into a Bayesian
statistical interpretation of the truncation errors underlying standard EFT
estimates~\cite{Furnstahl:2015rha}.  We applied this to construct the
probability distributions of the theoretical uncertainties quoted in
Eqs.~\eqref{eq:polsresults} and \eqref{eq:spinpols}; see
Ref.~\cite{latticepols} and references therein.  Since it does not employ
comparison with experiments but is based on information which is intrinsic to
the EFT expansion, the results fulfil a fundamental criterion of the
scientific method: falsifiability.

\begin{figure*}[!hbt]
\centering\includegraphics*[width=0.9\linewidth]{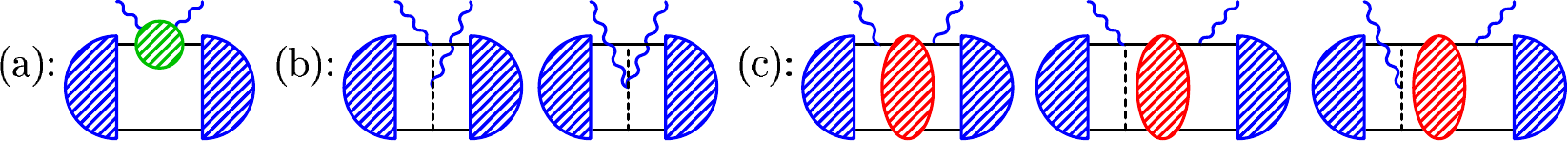}
\caption{\label{fig:contribs} Contributions to deuteron Compton scattering. Ellipse: $NN$ $S$-matrix:(a): single-nucleon; (b) photon
  coupling to charged exchange currents which bind the nucleus as dictated by
  chiral symmetry; (c) rescattering between emission and absorption restores
  the low-energy Thomson limit and guarantees current conservation.}
\end{figure*}

Such an uncertainty assessment is of course also vital for reliable
extractions of \emph{neutron} polarisabilities since one must
model-independently subtract nuclear binding effects from few-nucleon data.
Figure~\ref{fig:contribs} shows examples of the three classes of contributions
in few-nucleon systems.  Charged exchange currents and rescattering often
dominate over the targeted nucleonic structure contributions. An analysis of
Compton scattering therefore also provides indirect, non-trivial benchmarks as
to how accurately the chiral expansion accounts order-by-order for nuclear
binding and its mesonic contributions. For the deuteron, our results are
complete at $\calO(e^2\delta^3)$ or \NXLO{3} from the Thomson limit up to
about $120$~MeV, including the $\Delta(1232)$ degree of freedom~\cite{Gr12}.

\begin{figure*}[!htb]
  \centering\includegraphics*[width=0.47\linewidth]{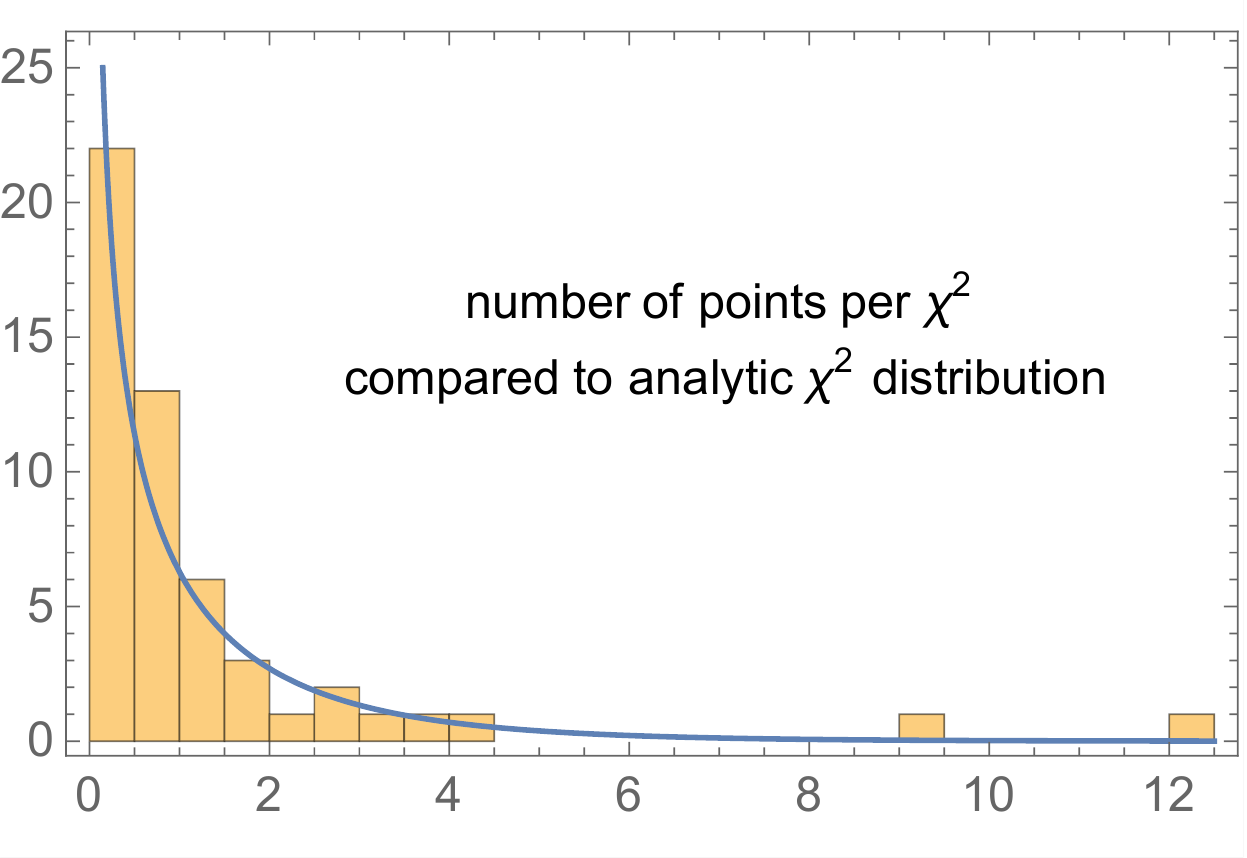}\hspace*{4ex}
\includegraphics*[width=0.47\linewidth]{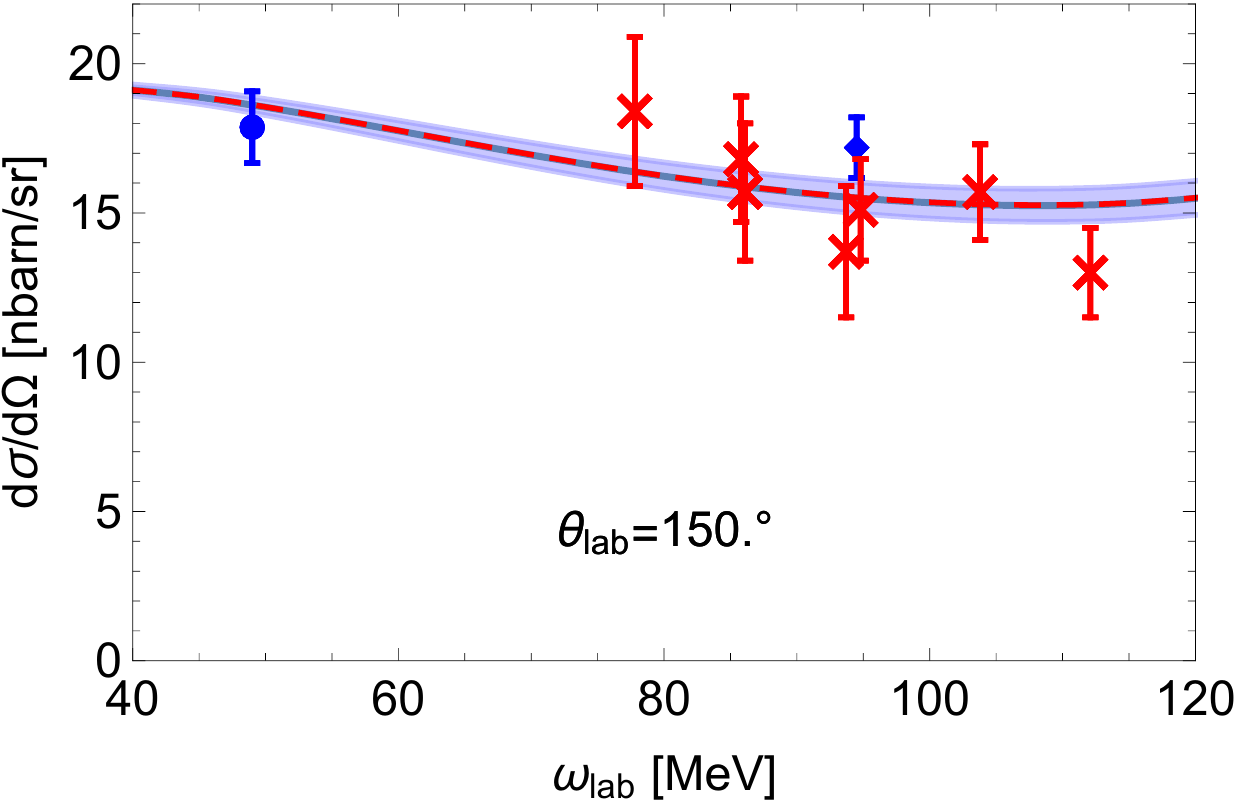}
\caption{\label{fig:pruning} \emph{Top}: Histogram of the number of deuteron Compton data
  with a given $\chi^2$, overlaid with the predictions of an ideal,
  statistically consistent set with $1$ degree of freedom ($1$ datum singled
  out, all others fixed). The two data with $\chi^2\ge9$ are pruned by
  statistical-likeliness criteria; including them does not have a significant
  impact on the neutron values. We add point-to-point and
  angle-dependent systematic errors in quadrature to the statistical error,
  and subsume overall systematic errors into a floating normalisation. The
  norm of each dataset floats by $\le5$\% and within the respective
  quoted normalisation errors of the data.
  \emph{Bottom}: Example of data and \ChiEFT result, with fit uncertainties
  (red: new MAXlab data); see Refs.~\cite{Myers:2014ace, Myers:2015aba} for
  details.}
\end{figure*}

Recently, 
$22$ 
points were added to the deuteron database by MAXlab~\cite{Myers:2014ace,
  Myers:2015aba}. This first new data in over a decade effectively doubled the
deuteron's world dataset. Our analysis shows that it is fully consistent with
and within the world dataset ($\chi^2=45.2$ for $44$ degrees of freedom), and
with the Baldin sum rule. Using the same Bayesian methods as in our
determinations of the proton and spin polarisabilities in
Eqs.~\eqref{eq:polsresults} and \eqref{eq:spinpols}, we assessed the
theoretical uncertainty as $\pm0.8
$. These data alone slashed the statistical error by $30$\%, with new values
adopted by the 2015 PDG. Just to illustrate the data quality, the $\chi^2$
distribution of the new world dataset agrees with the analytic expectation;
see Fig.~\ref{fig:pruning}. 
%

\section{Where We Want To Be}

\begin{figure*}[!htb]
\centering
  \includegraphics[width=0.47\linewidth,clip=]
{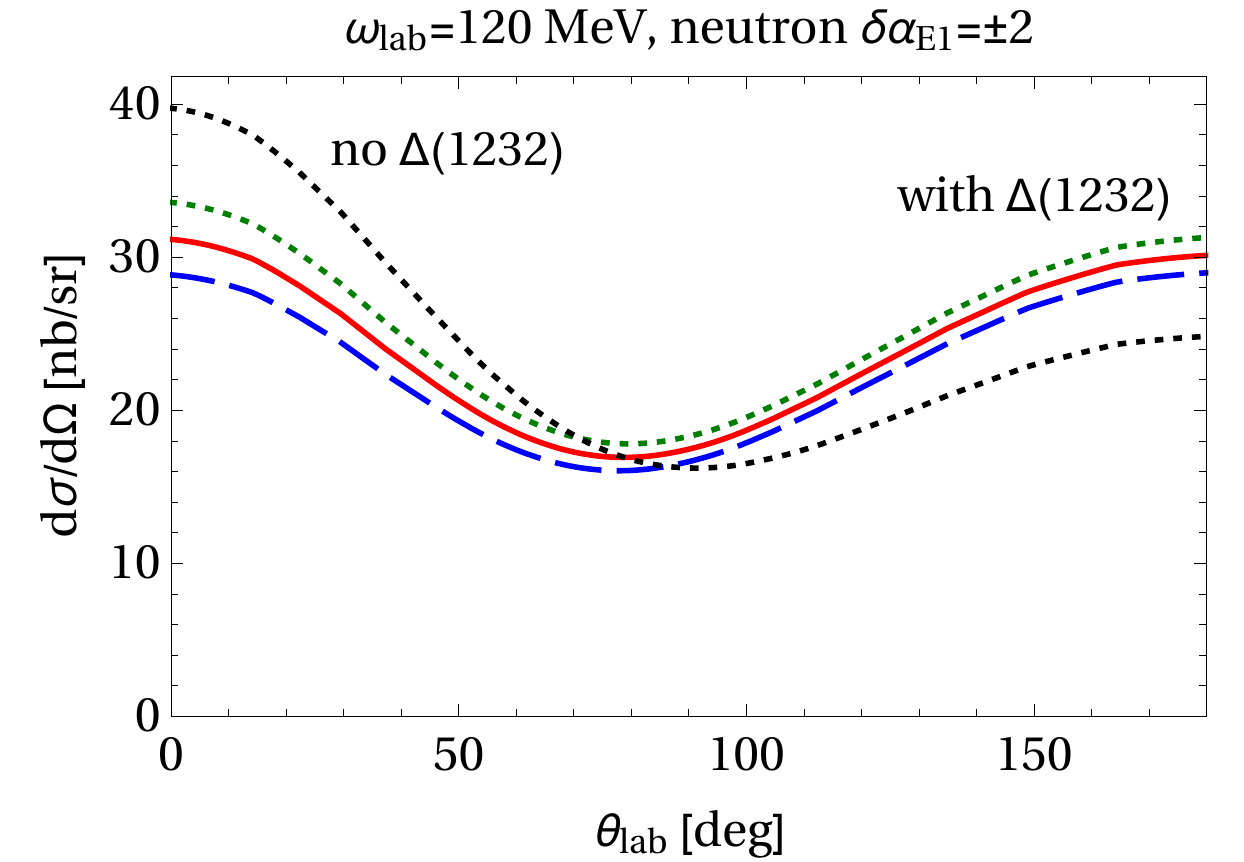}
\hspace{4ex}
\includegraphics
[width=0.47\linewidth]
{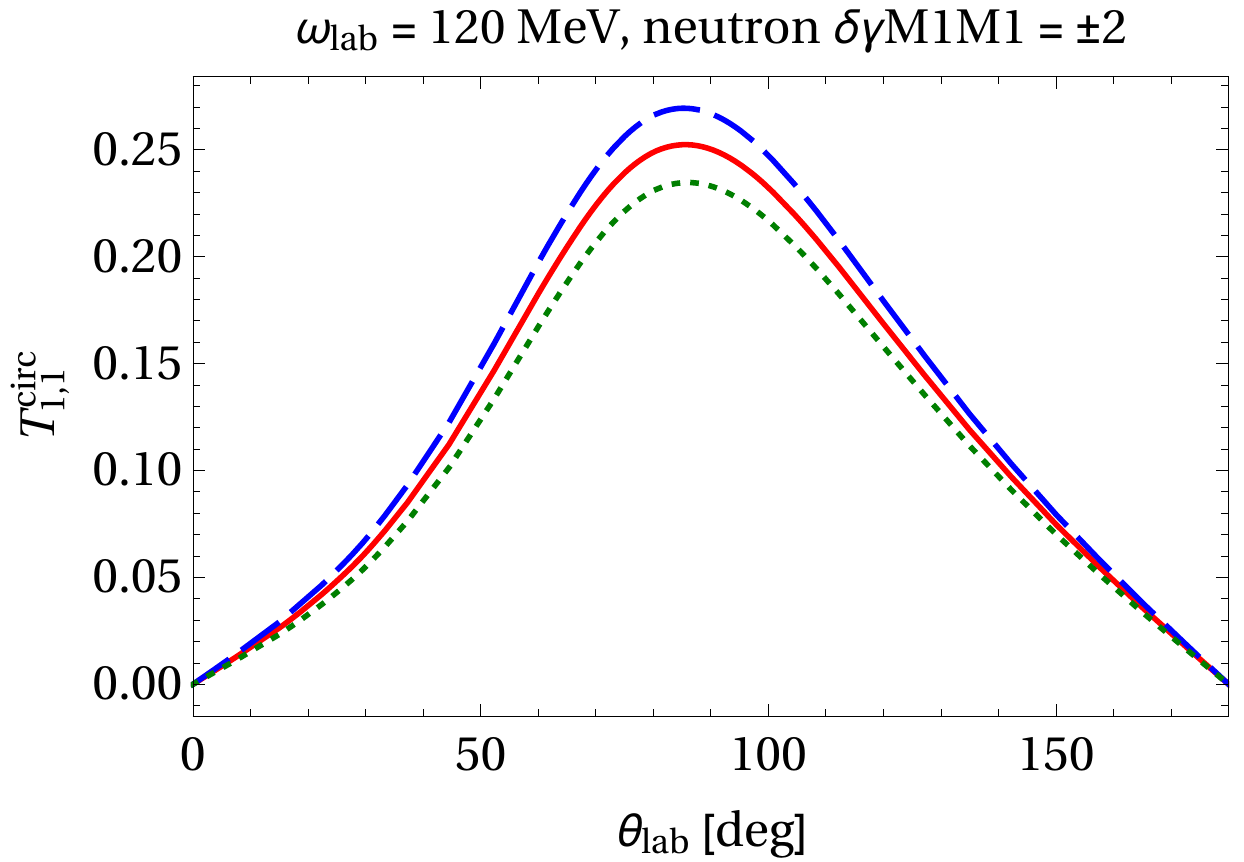}

\caption{\threeHe Compton scattering at
  $120\;\MeV$. \emph{Top:} \label{fig:3He}  
  with $\Delta(1232)$ (red solid), and
  without 
  (black dotted); blue dashed/green dotted: $\alphaen\pm2$~\cite{3He}.
  \emph{Bottom:} 
  \label{fig:3Hepolarised} Sensitivity of the asymmetry
  $T^\mathrm{circ}_{11}=-\sqrt{2}\Sigma_{2x}$ with $\Delta(1232)$ on
  $\gamma_{M1M1}^{(\mathrm{n})}$~\cite{3He}.}
\end{figure*}

The future lies in unpolarised, single-polarised and double-polarised
experiments of high accuracy, and in theoretical analyses with reproducible
systematic uncertainties. To understand the subtle differences of the pion
clouds around the proton and neutron induced by explicit chiral symmetry
breaking in QCD, we need to know the neutron polarisabilities with
uncertainties comparable to those of the proton -- Eq.~\eqref{eq:polsresults}
shows that this is mostly an issue of better data (and some theory work which
is under way). Therefore, MAMI, MAXlab and \HIGS aim for deuteron data with
statistical and systematic uncertainties of better than $5\%$, and plan
extensions to \threeHe and \fourHe
. In general, heavier nuclei are experimentally better to handle and provide
count rates which scale at least linearly with the target charge when photons
scatter incoherently from the protons, i.e.~for~$\omega\gtrsim100\;\MeV$.  But
a theoretical description of their energy levels with adequate accuracy is
involved. For the proton, amplitudes on the $\lesssim2\%$-level for
$\omega\lesssim\mpi$ and $\lesssim20\%$ around the $\Delta$ resonance are
available; for deuteron and \threeHe, we now extend descriptions with similar
accuracies into the Delta resonance region. Around \threeHe-\fourHe-${}^6$Li
may well be the ``sweet-spot'' between the needs and desires of theorists and
experimentalists.
For example, we updated the single-nucleon parts of the \threeHe code to the
same order $e^2\delta^3$ as in the deuteron.  Figure~\ref{fig:3He} shows that
excluding the energy-dependence due to the $\Delta(1232)$ can lead to false
signals in high-accuracy extractions of neutron polarisabilities. As in the
deuteron, the effect is increased at back-angles, but forward-rates are
suppressed; see Ref.~\cite{3He} for details.

\begin{figure}[!htb]
\centering
\includegraphics[width=1.0\linewidth]
{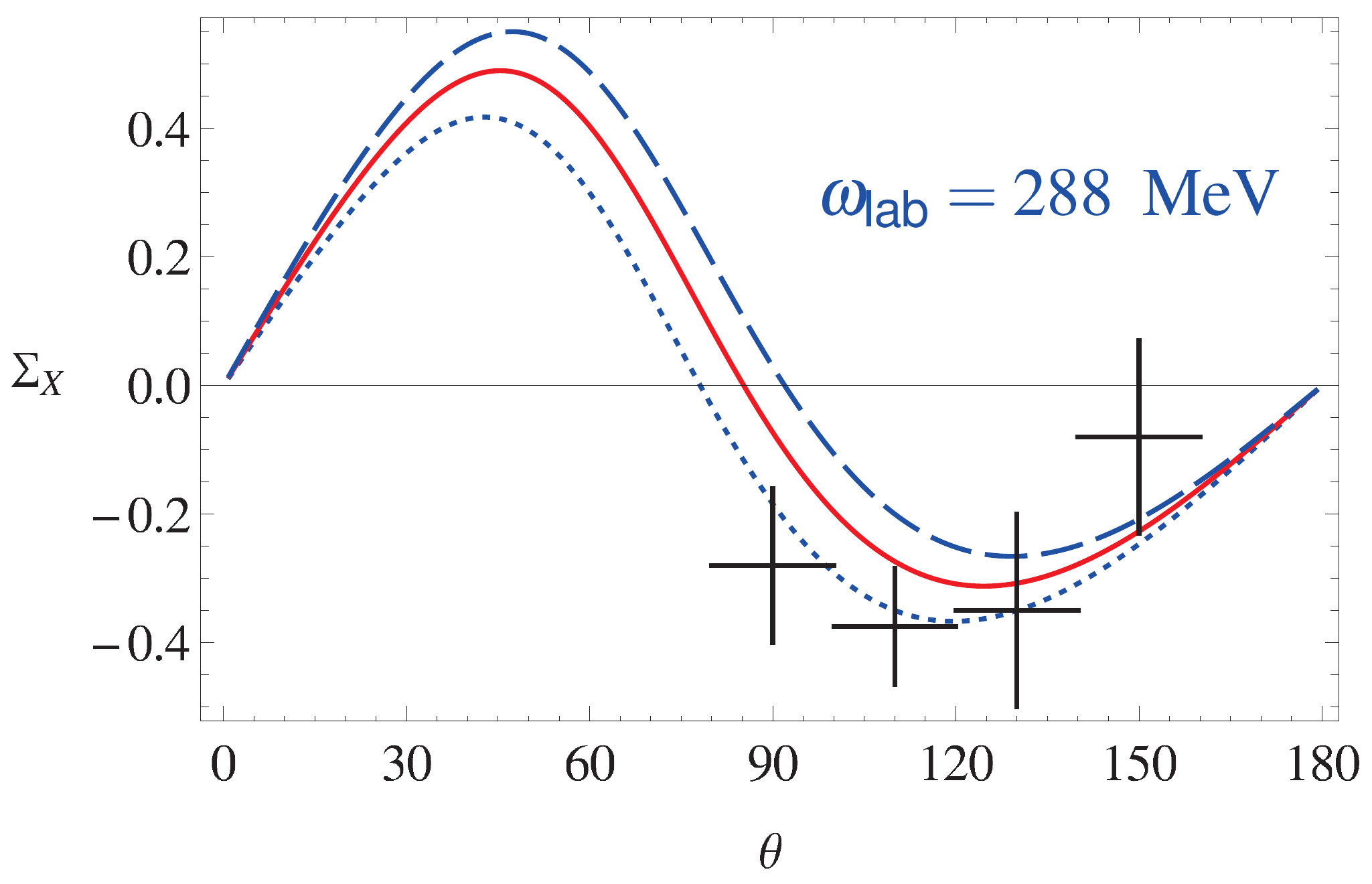}
\caption{\label{fig:Sigma2x} \ChiEFT prediction and MAMI
  data~\cite{Martel:2014pba} for the double-polarisation observable
  $\Sigma_{2x}$ on the proton. Solid: $\gammaee=-1.1$ (predicted);
  dashed/dotted: uncertainties of spin polarisabilities from
  Eq.~\protect\eqref{eq:spinpols} and Ref.~\protect\cite{latticepols} (other
  \ChiEFT truncation errors not included).}
\end{figure}

Since the four spin-polarisabilities for each nucleon -- which, as yet, are
hardly explored -- probe the spin-con\-stituents of the nucleon, they are a
top priority of experiment and theory alike. Sensitivity studies have been
performed in \ChiEFT variants with and without explicit $\Delta(1232)$; see
Fig.~\ref{fig:Sigma2x} and summary in~\cite[Sec.~6.1]{Gr12}. 
The MAMI experiment and extraction for the proton 
agrees well with our \ChiEFT findings, although both extraction and \ChiEFT
are at too high an energy to be really reliable, see Eq.~\eqref{eq:spinpols}
and Fig.~\ref{fig:Sigma2x}.

\begin{figure}[!htb]
\centering
\includegraphics
[width=\linewidth]
  {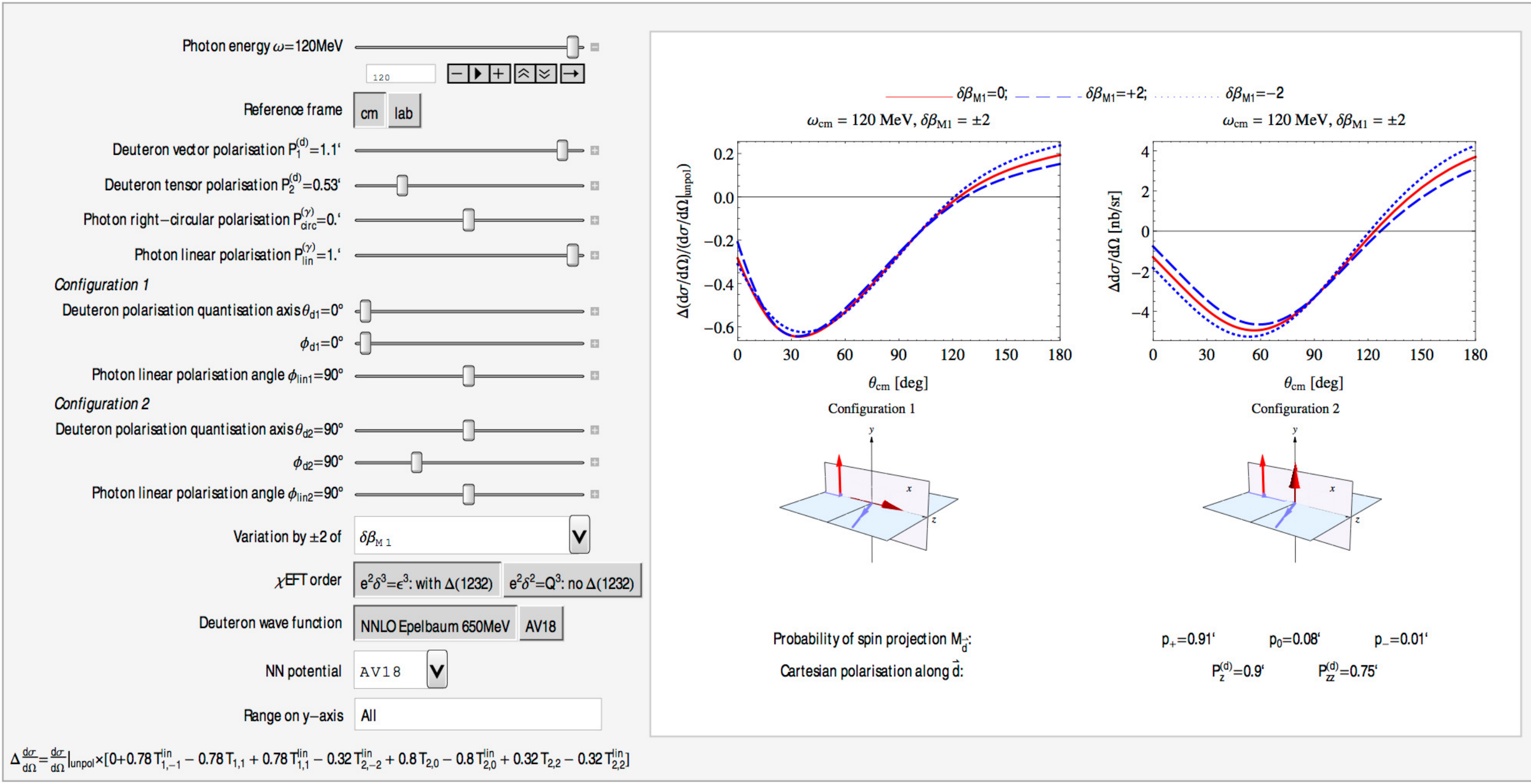}
  \caption{
    \label{fig:screenshot} \emph{Mathematica} screenshot for deuteron
    Compton scattering with arbitrary
    polarisations~\cite{Griesshammer:2013vga}.}
\end{figure}

Recently, the deuteron cross section and asymmetry with arbitrary photon and
target polarisations have also been parametrised via $18$ independent
observables~\cite{Griesshammer:2013vga}.  Particularly interesting are some
asymmetries which turn out to be sensitive to only one or two
polarisabilities. For spin polarisabilities with an error of
$\pm2\times10^{-4}\;\fm^4$, asymmetries should be measured with an accuracy of
$10^{-2}$ or so, with differential cross sections of a dozen nb/sr at
$100\;\MeV$ or a few dozen nb/sr at $250\;\MeV$. Relative to single-nucleon
Compton scattering, interference with the deuteron's $D$ wave and
pion-exchange current increases the sensitivity to the ``mixed'' spin
polarisabilities $\gammaem$ and $\gammame$.  A \emph{Mathematica} file for
$\omega<120\;\MeV$ is available from \texttt{hgrie@gwu.edu} (see screen-shot
in Fig.~\ref{fig:screenshot}), and more are being finalised for the proton and
\threeHe.

It is well-recognised that polarised \threeHe is (approximately) a polarised
neutron target. Corrections to this statement can be quantified using ab
initio wave functions calculated with \ChiEFT potentials.  \ChiEFT suggests
the photon beam asymmetry with a transervsely polarised target, $\Sigma_{2x}$,
is the cleanest observable in which to determine neutron spin
polarisabilities. It is affected by $\gamma_{M1M1}^{(\mathrm{n})}$ much more
strongly than it is by the scalar polarisabilities -- and unaffected by any
proton spin polarisabilities. Other neutron spin polarisabilities affect this
observable, but manifest a different angular dependence.  In fact, as \threeHe
is doubly charged and contains more nucleon pairs, interference of
polarisability effects with proton Thomson terms and meson-exchange currents
is larger than for the deuteron or proton. It may therefore be that neutron
spin polarisabilities can ultimately be determined more accurately than proton
ones---if the necessary target densities can be reached; see
Fig.~\ref{fig:3Hepolarised}.

\begin{figure*}[!htb]
  \centering\includegraphics*[height=0.33\linewidth]{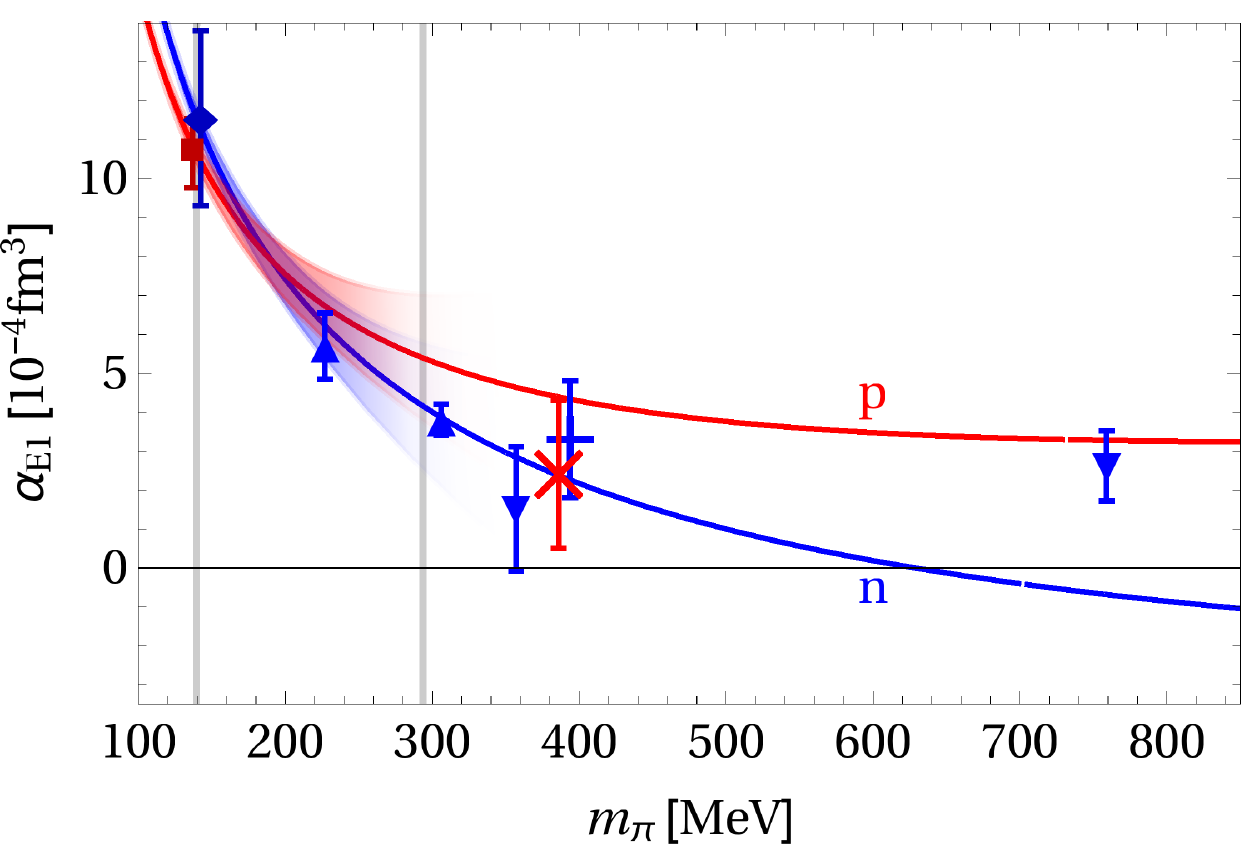}
  \hspace*{3ex}
\includegraphics*[height=0.33\linewidth]
{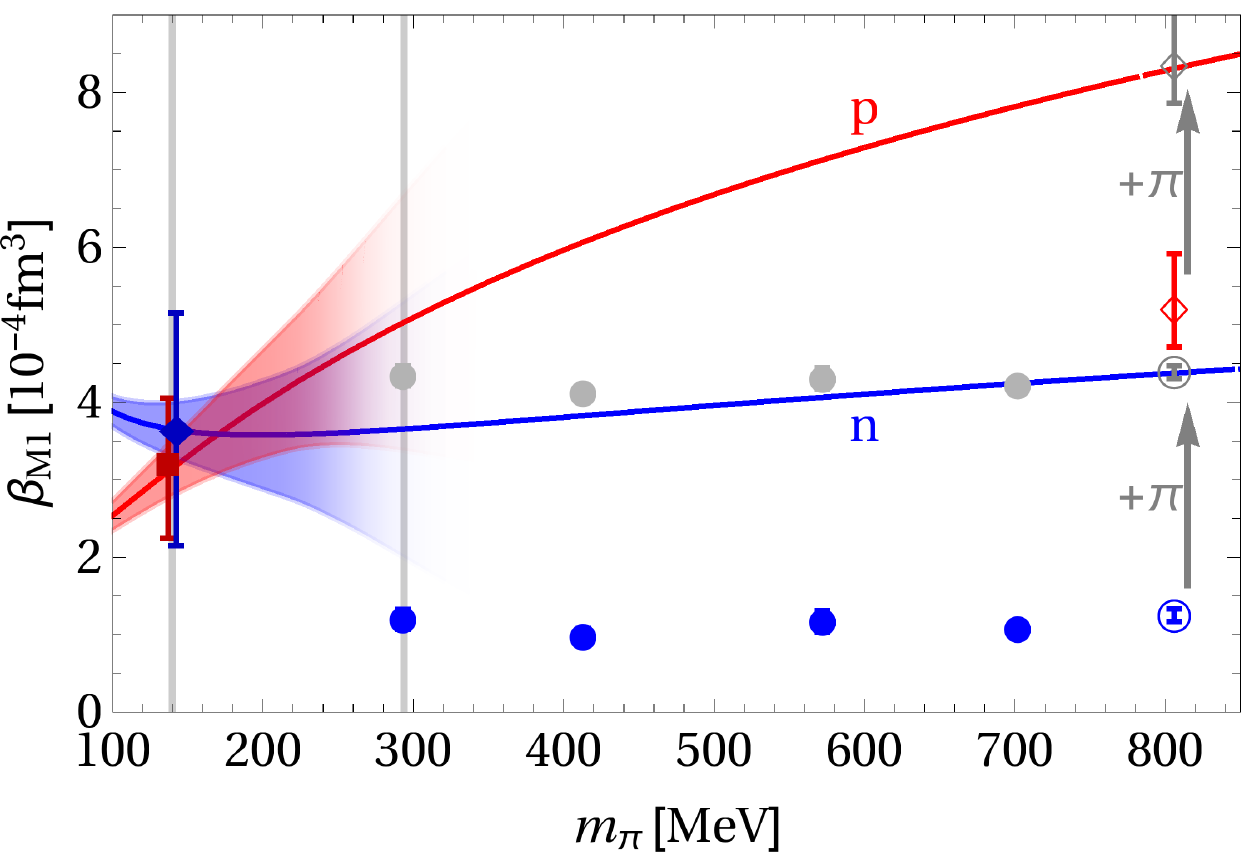}
\caption{\label{fig:lattice} Comparison of our \ChiEFT \emph{predictions} to
  lattice computations~\cite{latticepols}. Red/blue lines with red/blue error corridors: our
  \ChiEFT results. Corridors represent theoretical uncertainties in the regime
where \ChiEFT can be expected to converge and fade out as the uncertainty estimate becomes less
  reliable.  Error-bars at the
       physical point add statistical, theory and Baldin-sum-rule errors
       linearly, as applicable. \emph{Left}:  $\alphae$: 
       $\protect\textcolor{blue}{\blacktriangle}$ (neutron) Lujan et
       al.~\cite{Lujan:2014qga};
       $\protect\textcolor{red}{\times}$ (proton) and
       $\protect\textcolor{blue}{+}$ (neutron) Detmold et
       al.~\cite{Detmold:2010ts};
       $\protect\textcolor{blue}{\blacktriangledown}$ (neutron)
       Engelhardt/LHPC~\cite{Engelhardt:2007ub, Engelhardt:2010tm}.
\emph{Right}:
$\betam$: $\protect\textcolor{blue}{\bullet}$ (neutron) Hall et
al.~\cite{Hall:2013dva};
\protect\rotatebox{45}{$\protect\textcolor{red}{\scriptscriptstyle\square}$}
(proton) and $\protect\textcolor{blue}{\circ}$ (neutron)
NPLQCD~\cite{Chang:2015qxa}.  Gray ``ghost points'' found by shifting all
lattice results by $+\pi\times10^{-4}\;\fm^3$ [\emph{sic!}].}
\end{figure*}

Finally, \ChiEFT connects data with emerging lattice-QCD computations by
reliable extrapolations from numerically less costly, heavier pion masses
within the \ChiEFT regime to the physical point and circumvents a direct
lattice computation of Compton scattering -- that would be highly
nontrivial. Lattice computations, in turn, test to what extent \ChiEFT
adequately captures the $\mpi$-dependence of the low-energy dynamics, and may
predict short-distance (fit) parameters from QCD, as an alternative to
determining them experimentally. A particularly interesting \ChiEFT prediction
is a rather strong isovector component away from the physical point for both
$\alphae$ and $\betam$, arising from an intricate interplay of the chiral
physics of the pion cloud and short-distance effects.
The NPLQCD collaboration published
intriguing lattice results for the proton's and neutron's $\betam$ at
$\mpi=806\;\MeV$
~\cite{Chang:2015qxa}; see Fig.~\ref{fig:lattice} and~Ref.~\cite{latticepols
}.
The \emph{difference} $\betamp-\betamn$ is nearly identical to the chiral
result even well beyond the range in which \ChiEFT should be applicable. This
suggests that the experimental finding $\betamp \approx \betamn$ is something
of a coincidence. The agreement with lattice computations for $\alphae$ is
even better~\cite{latticepols}.

\begin{acknowledgments}
HWG cordially thanks the organisers, in particular for their scheduling
flexibility; and the audience for the strong reaction to a last-minute idea
how to start the talk. 
This work was supported in part by UK Science and Technology Facilities
Council grants ST/J000159/1 and ST/L005794/1 (JMcG), by the US Department of
Energy under contracts DE-FG02-93ER-40756 (DRP) and DE-FG02-95ER-40907 (HWG),
and by the Dean's Research Chair programme of the Columbian College of Arts
and Sciences of The George Washington University (HWG).
\end{acknowledgments}


\end{document}